\documentclass[12pt]{article}
\usepackage{latexsym}
\usepackage{amsmath,amssymb,mathrsfs}
\usepackage{type1cm}
\usepackage[dvips]{graphicx}

\begin{document}
\def\nospam{gravity.phys.nagoya-u.ac.jp}
\def\AT{@}

\title{The Role of Anisotropy in the Void Models without Dark Energy}

\author{Masayuki Tanimoto\thanks{E-mail: \texttt{tanimoto\AT\nospam}}, 
Yasusada Nambu\thanks{E-mail: \texttt{nambu\AT\nospam}} \ and 
Kazuhiro Iwata
 \vspace{.5em}
\\
\textit{Department of Physics, Graduate School of Science,} \\
\textit{Nagoya University, Chikusa, Nagoya 464-8602, Japan}
}
\maketitle

\begin{abstract}
  Void models provide a possible explanation of the ``accelerated
  expansion'' of the Universe without dark energy. To make the
  conventional void models more realistic, we allow the void, an
  underdense region around us, to be anisotropic and consider an
  average of the distance-redshift relations over the solid angle
  subtended at the observer. We first show that after taking the
  average of a form of the optical scalar equation (distance
  equation), the effective distance equation we obtain coincides with
  the one for the Lema{\^i}tre-Tolman-Bondi universe with a
  Dyer-Roeder-like extension. We then numerically solve the equation
  to compare with observational data of Type Ia supernovae. We find
  that anisotropy allows smaller size of void and larger $\Omega_m$.
\end{abstract}

\section{Introduction}
In recent years, reliable distance-redshift relations up to $z\sim 1$
have been revealed by observations of Type Ia supernovae
\cite{Riess:1998cb,Perlmutter:1998np}. Comparisons of the data
obtained from these observations with the predictions from
Friedmann-Lema\^{i}tre (FL) homogeneous and isotropic universe models
suggest that the Universe is in a phase of accelerated expansion, and
there must exist unidentified dark energy that causes the
acceleration. The nature of dark energy has been investigated by many
researchers \cite{copeland-2006-15}.

On the other hand, the possibilities have been discussed that the
distance-redshift relations provided by the observations of Type Ia
supernovae may be reproduced by taking account of the inhomogeneities
in the matter distribution and geometry without dark energy. One of
such models is the so-called void model, in which existence of a large
scale underdense region (void) is assumed. Actually, an inhomogeneous
distribution of matter of scale $200\sim 300/h$ Mpc (the Hubble
constant is $100h$ km/s/Mpc) was suggested by several
authors~\cite{Marinoni:1999wg} in  galactic surveys, and a void on
scales of about $250/h$ Mpc was found by Blanton \textit{et
  al}.~\cite{Blanton:2000ns} in the SDSS commissioning data. These
results suggest that we might live in a local void with a radius of
$200\sim 300/h$Mpc.  Motivated by these observations,
Tomita~\cite{Tomita:1999rw,Tomita:2001gh,Tomita:2000jj} sought to
reproduce the distance-redshift relations using a model in which a
homogeneous spherical region around us is less dense than the
background.

More recently, Alnes \textit{et al}.~\cite{Alnes:2005rw} (also,
\cite{Alexander:2007xx,GarciaBellido:2008nz}) showed that there exists
a model that provides a good fit to the Type Ia supernova observations
(as well as other kinds), assuming that the outside of a spherical
region is homogeneous and isotropic but the inside of it is described
by the Lemaitre-Tolman-Bondi (LTB) spherically symmetric dust
solution~\cite{Lemaitre:1933gd,Tolman:1934za,Bondi:1947av}. Yoo
\textit{et al}.~\cite{Yoo:2008su} showed that if the arbitrary
functions in the LTB solution are determined so that the
distance-redshift relation in the LTB spacetime coincides with that of
the $\Lambda $CDM model, the resulting model has a void structure on
scales of a few Gpc.

The essential mechanism for these void models to be able to reproduce
the distance-redshift relations provided by the observations is that
the local Hubble expansion rate monotonically decreases as the
distance increases so that the incoming light rays experience
effective increase in the expansion rate.

In reality, however, the matter distribution in the Universe is not
isotropic, and the distance-redshift relation inferred by the
observational data should be considered as an average over all the
directions of sight. This point seems to have been largely ignored in
the literature. To take into account the effect caused by these
anisotropies, we in this paper start with the optical scalar equation
in general inhomogeneous and anisotropic spacetimes, and then take an
average of a form of this equation (distance equation) over the
spheres on the past light cone.  We find that after some estimates,
the resulting equation coincides with the one in the LTB universe with
the matter density $\rho_{\text{LTB}}$ formally replaced by $\alpha(z)
\rho_{\text{LTB}}$, where $\alpha(z)$ is a certain function of $z$. In
other words, the equation coincides with the one obtained as a
Dyer-Roeder-like~\cite{dyer:1972,dyer:1973} extension of the distance
equation in the LTB universe. With this effective equation, we
numerically investigate light properties of void models including
effects of anisotropy.

Mattsson~\cite{Mattsson:2007tj} has recently discussed a similar
generalization of the Dyer-Roeder equation, arguing principally that
the original Dyer-Roeder extension did not take into account spatial
inhomogeneities in expansion rate. To incorporate these
inhomogeneities, he replaced the Hubble expansion rate $H_\text{FL}$
in the Dyer-Roeder equation with $\beta_\text{M}(z) H_\text{FL}$, with
$\beta_\text{M}(z)$ being an arbitrary function of $z$. We do not have
to artificially introduce a phenomenological function like
$\beta_\text{M}(z)$, since we start with a general inhomogeneous and
anisotropic setting, which naturally leads to an equation with the
effect of inhomogeneities in both matter density and expansion
rate. Moreover, the inhomogeneity and anisotropy we consider is
supposed to be made by smooth distribution of dark matter, rather than
the spiky clumpiness by galaxies as assumed in
Mattsson~\cite{Mattsson:2007tj} and in Dyer and
Roeder~\cite{dyer:1972,dyer:1973}.

This paper is organized as follows. In section \ref{main}, we derive
our effective distance equation by means of averaging. In section
\ref{numsol}, we consider a specific void model and numerically solve
the effective distance equation to search the parameter space for the
best values that are most likely to explain observational data.
Section \ref{conclusion} is devoted to conclusion and some remarks.
Appendix \ref{originalD-R} provides a review for the original
Dyer-Roeder extension. Throughout this paper, we use $c=1$ as the
speed of light.

\section{The effective distance equation for inhomogeneious and
 anisotropic spacetimes}
\label{main}

In the first subsection below, we transform the optical scalar
equation to derive an equation for the angular-diameter distance $D_A$
in an inhomogeneous and anisotropic dust universe. We then specialize
the equation to the LTB spherically symmetric dust universe case for
later use. In the second subsection, we take a spherical average of
the general distance equation to make the equation suitable for
comparisons with observational data.

\subsection{Distance equation and its specialization in the LTB spacetime} 
\label{LTB}

We start with the optical scalar equation~\cite{Sachs:1961zz}, which
determines the variations of the cross sectional area of a null
geodesic congruence. Let us assume that the null geodesics in the
congruence arrive at the observer through paths that are sufficiently
far from strong inhomogeneities such as galaxies. This allows us to
ignore the gravitational lensing effect, and in turn, the shear and
rotation of the congruence. The optical scalar equation in this case
becomes
\begin{equation}
\frac{2}{\sqrt{A}}\frac{d^2\sqrt{A} }{d\lambda^2 }
+R_{\mu \nu }k^\mu  k^\nu  =0,  \label{eq:sachs}
\end{equation}
where $A$ is the cross sectional area of the congruence, $k^\mu$ the
null geodesic generator, $\lambda $ the affine parameter, and $R_{\mu
  \nu }$ the Ricci tensor. On the other hand, Einstein's equation
\begin{equation}
R_{\mu \nu }-\frac{1}{2}Rg_{\mu \nu }=8\pi GT_{\mu \nu },  
\label{eq:einstein}
\end{equation}
where $T_{\mu \nu }$ is the energy momentum tensor, implies
\begin{equation}
\label{eq:R2T}
R_{\mu \nu }k^\mu  k^\nu =8\pi GT_{\mu \nu }k^\mu  k^\nu, 
\end{equation}
since $k^\mu$ is null. With the relation $D_A\propto\sqrt{A}$ for the
angular diameter distance $D_A$, we obtain
\begin{equation}
\frac{2}{D_A}\frac{d^2D_A }{d\lambda^2 }
+8\pi GT_{\mu \nu }k^\mu  k^\nu  =0 \label{eq:Ray}
\end{equation}
from Eqs.(\ref{eq:sachs}) and (\ref{eq:R2T}).  The only energy content
we consider is dust, so that $T_{\mu \nu }=\rho u^\mu u^\nu$, where
$\rho$ is the energy density for dust, $u^\mu$ the 4-velocity of dust.
The frequency of the light observed by a comoving observer at a
given $\lambda$ with 4-velocity $u^\mu$ is given by
$\omega(\lambda)=-u_\mu k^\mu$ (e.g., \cite{Wald:text}). Therefore the
second term in the left hand side of Eq.(\ref{eq:Ray}) yields
\begin{align}
T_{\mu \nu }k^\mu  k^\nu &=\omega(\lambda )^2 \rho \notag \\
                         &=\omega(0)^2 (1+z)^2\rho, \label{eq:dust}
\end{align}
where we have used the definition of redshift $z$, $z\equiv
\frac{\omega(\lambda )}{\omega(0)}-1$. Thus, Eq.(\ref{eq:Ray}) can be
written as
\begin{equation}
\mathcal{L}^2 D_A
+4\pi G\rho D_A=0, \label{eq:generalD-R}
\end{equation}
where $\mathcal{L}^2$ is the second order linear derivative operator
defined by
\begin{equation}
  \label{eq:L^2}
  \mathcal{L}^2\equiv \frac{1}{\omega^2(0)(1+z)^2}\frac{d^2}{d\lambda ^2}.
\end{equation}
We stress that this equation is valid for any dust universe, as long
as the gravitational lensing effect along the geodesics we consider
can be ignored. Note also that solving this equation requires another
equation (geodesic equation) to relate $\lambda$ with $z$, for given
$\rho$.

Next, we specialize the last equation to the LTB spherically symmetric
spacetime case with the observer at the symmetry center. The essential
part of our task is to convert $d/d\lambda$ to $d/dz$ for the case.

The LTB metric is given by
\begin{align}
ds^2 &=-dt^2+\frac{(R'(t,r))^2}{1+2E(r)}dr^2+R(t,r)^2d\Omega ^2, 
\end{align}
where the `energy function' $E(r)$ is an arbitrary function concerning
curvature. The circumferential radius $R(t,r)$ is the solution for the
generalized Friedmann equation
\begin{equation}
\frac{1}{2}\dot R^2(t,r)-\frac{M(r)}{R(t,r)}=E(r), \label{eq:Einstein}
\end{equation}
where the `mass function' $M(r)$ is another arbitrary function, and
$\dot{ }$ and ${}'$ denote the derivatives with respect to $t$ and
$r$, respectively. 

Using the analytic function $\mathcal{S}(x)$ introduced in
Ref.\cite{Tanimoto:2007dq}, the solution $R$ can be represented simply
as
\begin{equation}
R(t,r)=(6M(r))^{1/3}(t-t_B(r))^{2/3}
\mathcal{S}\left(-2E(r)\left(\frac{t-t_B(r)}{6M(r)}\right)^{2/3}\right),
\label{eq:R-S}
\end{equation}
where the `Big-Bang function' $t_B(r)$ is a third arbitrary
function. The function $\mathcal{S}(x)$ is the unique solution of the
first order ordinary differential equation
\begin{equation}
  \label{eq:SODE1}
  \frac43(\mathcal{S}(x)+x \frac{d\mathcal{S}(x)}{dx})^2
  +3x-\frac1{\mathcal{S}(x)}=0
\end{equation}
that intersects the $x=0$ axis transversely
\cite{Tanimoto:2007dq}. This function is non-negative,
monotonically decreasing, and defined for $x\leq (\pi /3)^{2/3}$.

As for the dust matter, the 4-velocity is given by $u^\mu=(1,0,0,0)$,
and the energy density by
\begin{equation}
  \label{eq:rhoLTB}
  4\pi \rho_{\text{LTB}} (t,r)=\frac{M'(r)}{R'(t,r)R^2(t,r)}.
\end{equation}

Now, suppose that the light we observe progresses along a path
$x^\mu(\lambda)$ for decreasing $\lambda$ with $\lambda=0$
corresponding to the observer, so that $k^\mu=-dx^\mu/d\lambda$. Then,
we obtain
\begin{align}
\frac{d}{d\lambda } 
 &=\frac{dz}{d\lambda}\frac{d}{dz}
 =\frac{1}{\omega(0)}\frac{d\omega(\lambda)}{d\lambda}\frac{d}{dz}\notag \\
 &=\frac{-1}{\omega(0)}\left(\frac{d^2t}{d\lambda^2} \right)\frac{d}{dz},
 \label{eq:lam-z}
\end{align}
since $\omega(\lambda)=k^0(\lambda)=-dt/d\lambda$. On the other hand,
the radial geodesic equation for $t(\lambda)$ and the null condition,
respectively, give us
\begin{align}
\frac{d^2t}{d\lambda ^2} 
&=\frac{-R'\dot{R'}}{1+2 E}\left(\frac{dr}{d\lambda } \right)^2, &
-\left(\frac{dt}{d\lambda } \right)^2 
+\frac{R'^2}{1+2 E}\left(\frac{dr}{d\lambda } \right)^2 &=0.
\end{align} 
Using these equations, we immediately have
\begin{align}
\frac{d}{d\lambda } 
&=\frac{1}{\omega(0)}\frac{\dot{R'}}{R'}\left(
  \frac{dt}{d\lambda } \right)^2\frac{d}{dz} \notag \\
  &=\omega(0)\frac{\dot{R'}}{R'}(1+z)^2\frac{d}{dz} \notag \\
  &=
  \omega(0)H_{\text{LTB}}(z)(1+z)^2\frac{d}{dz}, \label{eq:l-zLTB}
\end{align}
where we have defined the effective Hubble function $H_{\text{LTB}}$,
\begin{equation}
  \label{eq:HLTB}
  H_{\text{LTB}}= \frac{\dot{R'}}{R'}.
\end{equation}

Consequently, Eq.(\ref{eq:generalD-R}) yields
\begin{equation}
  \mathcal{L}^2_{\text{LTB}} D^{(\text{LTB})}_A(z)
  +4\pi G\rho_{\text{LTB}}(z) D^{(\text{LTB})}_A(z)=0, 
  \label{eq:D-RLTB}
\end{equation}
where  
\begin{equation}
\mathcal{L}^2_{\text{LTB}} \equiv 
H_{\text{LTB}}(z)\frac{d}{dz}\left[ 
  H_{\text{LTB}}(z)(1+z)^2\frac{d}{dz} \right],
\end{equation}
and $D^{(\text{LTB})}_A$ represents the angular-diameter distance in
the LTB universe. Note that this equation is the same as the one for
the FL universe (See Eq.(\ref{eq:D-RFLRWDA}) in appendix
\ref{originalD-R}) with the Hubble function replaced by
Eq.(\ref{eq:HLTB}). (This is the reason we call $H_\text{LTB}$ the
effective Hubble function.) Note also that the effective Hubble
function $H_{\text{LTB}}(z)$ coincides with the so-called longitudinal
Hubble function $H_{\parallel }\equiv {\dot{R'}}/{R'}$; in contrast,
the perpendicular Hubble function $H_\perp\equiv\dot R/R$ plays no
role in determining the distance.

\subsection{Averaging of anisotropies} \label{general}

In this subsection, we take a spherical average of the distance
equation (\ref{eq:generalD-R}) and compare it with the one for the
isotropic spacetime Eq.(\ref{eq:D-RLTB}).

For our purpose, the most convenient is to take a particular
coordinate system that is based on a foliation of spacetime by past
light cones of a timelike curve. We employ a spherical null coordinate
system $(\tau,z,\theta,\phi)$ such that $\tau=\text{constant}$ on each
light cone, that $z=0$ corresponds to the vertex of each light cone,
that the timelike curve the vertices $z=0$ comprise coincides with the
orbit of matter flow there, and that the angular coordinates
$(\theta,\phi)$ are constant along each null geodesic that reaches the
vertex $z=0$. (This choice of angular coordinates is possible at least
in the domain where the null geodesics do not intersect each other.)
The radial null coordinate $z$ is chosen to give the redshift along
each null geodesic labeled by $(\tau,\theta,\phi)$. Our observation is
supposed to be done at the coordinate center $z=0$ at a certain
instant $\tau=\tau_0$.

Let $D_A(z\, ;\theta,\phi)$ be the angular-diameter distance to an
object observed in the direction $(\theta,\phi)$ with redshift
$z$. This is a function of $z$, $\theta$, and $\phi$, but we will
often regard this as a function of $z$ with $\theta$ and $\phi$ being
parameters, since we mostly consider its variations along the null
geodesic, in which situation $\theta$ and $\phi$ are constant. We will
therefore use $d/dz$ instead of $\partial/\partial z$ when it acts on
$D_A(z\, ;\theta,\phi)$. The same rule will be applied to other
functions as well.

Let us define the spherical average of a function $f(z\,;\theta,\phi)$
by
\begin{equation}
  \label{eq:sav}
\bar f(z)\equiv \langle f(z;\theta ,\phi)\rangle\equiv
\frac{1}{4\pi }\int f(z;\theta ,\phi )d\Omega,
\end{equation}
where $d\Omega\equiv\sin\theta d\theta d\phi$. Then, the spherical average
of Eq.(\ref{eq:generalD-R}) can be written as
\begin{equation}
\langle \mathcal{L}^2 D_A(z;\theta ,\phi ) \rangle
+4\pi G \langle \rho(z;\theta ,\phi ) 
D_A(z;\theta ,\phi )\rangle =0. 
\label{eq:avSachs}
\end{equation}
Here, the derivative operator $\mathcal{L}^2$ becomes under our
coordinates 
\begin{equation}
  \begin{split}
\mathcal{L}^2 &=
\frac{1}{\omega^2(0)(1+z)^2}
\left( \frac{dz}{d\lambda}\frac{d}{d z} \right)^2 \\
&= H(z\, ;\theta,\phi)\frac{d}{d z}\left[
    H(z\, ;\theta,\phi)(1+z)^2\frac{d}{d z}
  \right],
  \end{split}
\label{eq:Dz}
\end{equation}
where the general Hubble function $H(z\, ;\theta,\phi)$ is defined by
\begin{equation}
  \label{eq:genH}
  H(z\, ;\theta,\phi)\equiv \frac1{\omega(0)(1+z)^2}\frac{dz}{d\lambda}.
\end{equation}

Note that in the averaged equation (\ref{eq:avSachs}), each term in
the left hand side depends only on $z$, as in the LTB equation
(\ref{eq:D-RLTB}). We wish to estimate how much different the averaged
distance function $\bar D_A(z)=\langle D_A(z\, ;\theta,\phi)\rangle$
is from an LTB distance function $D^\text{(LTB)}_A(z)$. To this, we
match an LTB solution with the given inhomogeneous and anisotropic
universe according to the following correspondence:
\begin{equation}
  \label{eq:cor}
  \bar\rho(z)=\rho_\text{LTB}(z),\quad
  \bar H(z)=H_\text{LTB}(z).
\end{equation}
Since the LTB solution possesses two arbitrary physical functions (one
of the three arbitrary functions corresponds to gauge), it is natural
to expect that these two conditions are satisfied for a suitable
choice of these functions. (Strictly speaking, asserting this leads to
a variation of the inverse problem for the LTB solution, which we
consider out of our scope. For a discussion of the standard version of
the inverse problem, see, e.g., \cite{Mustapha:1998jb}.)

\def\LLL{\mathcal{L}_{\text{LTB}}^2}
\def\LL{\mathcal{L}^2}
\def\barDA{\bar D_A}
\def\angl#1{\langle #1 \rangle}
\def\anglgg#1{\bigg\langle #1 \bigg\rangle}
\def\HLTB{H_{\text{LTB}}}

Keeping this correspondence in mind, we formally rewrite equation
(\ref{eq:avSachs}) in the form
\begin{equation}
\langle \mathcal{L}^2 D_A \rangle (z)
+4\pi G \alpha (z)\rho_\text{LTB}(z) \bar D_A(z) =0,
\label{eq:abD-R}
\end{equation}
where
\begin{align}
\alpha (z) &\equiv 
\frac{\langle\rho(z;\theta ,\phi) D_A(z;\theta ,\phi)\rangle}{
  \rho_\text{LTB}(z) \bar D_A(z)}.
\end{align}

To evaluate the average $\langle \mathcal{L}^2 D_A \rangle$ and
the ratio function $\alpha(z)$, let us decompose physical functions as
\begin{align}
\rho(z;\theta ,\phi) &= \rho_{LTB}(z)[1+\delta_\rho(z;\theta ,\phi)], \\
H(z;\theta ,\phi) &= H_\text{LTB}(z)[1+\delta_H(z;\theta ,\phi)], \\
D_A(z;\theta ,\phi) &= \bar D_A(z)[1+\delta_D(z;\theta ,\phi)],
\end{align}
where $\langle \delta_* \rangle=0 \, (*=\rho, H, \text{or } D)$.  To
calculate the quantity $\mathcal{L}^2 D_A$, it is most
convenient to use the following identity that is confirmed by a
straightforward computation;
\begin{equation}
  \LL=(1+\delta_H)^2\LLL + \HLTB^2(1+z)^2[\delta_H'+\frac12
  (\delta_H^2)']\frac{d}{dz}.
\end{equation}
Here, a prime stands for $d/dz$. Using this formula we can calculate
\begin{equation}
  \begin{split}
    \mathcal{L}^2 D_A = & \LL \barDA(1+\delta_D) \\
    = & (\LL\barDA)(1+\delta_D)+\barDA\LL(1+\delta_D)+
    2\HLTB^2(1+\delta_H)^2(1+z)^2\barDA'\delta_D' \\
    = & (1+\delta_H)^2[(1+\delta_D)\LLL\barDA+\barDA\LLL\delta_D] \\
    & + \HLTB^2(1+z)^2(\delta_H'+\frac12(\delta_H^2)')[
    (1+\delta_D)\barDA'+\barDA\delta_D'] \\
    & + 2\HLTB^2 (1+z)^2 (1+\delta_H)^2\barDA'\delta_D'.
  \end{split}
\end{equation}
(To obtain the second line, Leibniz's rule was applied.)
Taking average of this equation leads to an expression for $\langle
\mathcal{L}^2 D_A \rangle$, which we, for convenience, write in three
parts depending on order in $\delta_*$;
\begin{equation}
    \langle \LL D_A \rangle =
    \LLL\barDA+O(2)+O(3),
\end{equation}
where $O(2)$ and $O(3)$ are, respectively, the second and third order
term. Brief calculations show that
\begin{equation}
  \begin{split}
    O(2) = & \,
    (\angl{\delta_H^2}+2\angl{\delta_H\delta_D})\LLL\barDA \\
    & +
    \HLTB^2(1+z)^2(\angl{\delta_H\delta_H'}+
    4\angl{\delta_H\delta_D'}+\angl{\delta_H'\delta_D})\barDA' \\
    & +
    \bigg( 2\angl{\delta_H\LLL\delta_D})
    +\HLTB^2(1+z)^2\angl{\delta_H'\delta_D'} \bigg) \barDA,
  \end{split}
\end{equation}
and
\begin{equation}
  \begin{split}
    O(3) = & \,
    \angl{\delta_H^2\delta_D} \LLL\barDA \\
    & +
    \HLTB^2(1+z)^2(\angl{\delta_H\delta_H'\delta_D}+
    2\angl{\delta_H^2\delta_D'})\barDA' \\
    & +
    \bigg( \angl{\delta_H^2\LLL\delta_D}
    +\HLTB^2(1+z)^2\angl{\delta_H\delta_H'\delta_D'} \bigg) \barDA.
  \end{split}
\end{equation}

Since the perturbation of distance is mainly caused by a perturbation
of the expansion rate, we may have $\delta_D\sim \delta_H$ as a rough
estimate. Simplifying the second order term according to this
approximation and omitting the third order term, we have
\begin{equation}
  \begin{split}
    \langle \LL D_A \rangle \simeq & \,
    (1+3 \angl{\delta_H^2})\LLL\barDA +
    3\HLTB^2(1+z)^2 \angl{\delta_H^2}'\barDA' \\
    & +
    \bigg( 2\angl{\delta_H\LLL\delta_H}
    +\HLTB^2(1+z)^2\angl{\delta_H'^2} \bigg) \barDA.
  \end{split}
\end{equation}

On the other hand, the function $\alpha(z)$ can be simply calculated as
\begin{align}
\alpha(z) &=
\frac{1}{\rho_{LTB}(z) \bar D_A(z)} \big\langle
\rho_{LTB}(z)[1+\delta_\rho(z;\theta ,\phi)] \,
\bar D_A(z)[1+\delta_D(z;\theta ,\phi)]  \big\rangle  \notag \\
&=1+\langle \delta_\rho \delta_D \rangle. 
\label{eq:LTBalpha}
\end{align}
The fact that the emptier the region the light we
observe travels, the longer the distance from the light source
implies that $\delta_\rho$ and $\delta_D$ have an opposite sign each
other. Therefore we have
\begin{equation}
  \label{eq:alphasign}
  \alpha(z) \leq 1
\end{equation}
with equality being satisfied only in the LTB isotropic universe.

Furthermore, according to the CMB observation, the perturbation of the
expansion rate is about  $|\delta_H|\equiv\sqrt{\angl{\delta_H^2}}\sim
0.1$~\cite{Wang:1997tp} on scales of $z\sim 0.1$, which implies that
the second order terms in $\langle \LL D_A \rangle$ are much smaller
than the second order term in $4\pi G\alpha(z)\rho_\text{LTB}\barDA$
if $|\delta_\rho|\sim 1$. We therefore neglect the second order terms
in $\langle \LL D_A \rangle$ and simply have
\begin{equation}
  \langle \LL D_A \rangle \simeq \LLL\barDA.
\end{equation}
We also note that if $|\delta_D|\sim |\delta_H|\sim 0.1$ and
$|\delta_\rho|\sim 1$, then Schwarz's inequality implies
$|\angl{\delta_\rho \delta_D}|\leq |\delta_\rho| |\delta_D|\sim 0.1$,
or
\begin{equation}
  \alpha(z)\gtrsim 0.9.
\end{equation}

To summarize, the distance equation (\ref{eq:abD-R})
takes the form
\begin{equation}
  \mathcal{L}^2_\text{LTB} \bar D_A(z)
+4\pi G \alpha (z)\rho_{\text{LTB}}(z) \bar D_A(z) =0,
\label{eq:alD-R}
\end{equation}
where $\alpha(z)$ is a function in the range
\begin{equation}
  0.9 \lesssim \alpha(z) \leq 1.
\end{equation}
Equation (\ref{eq:alD-R}) is the effective distance equation the
averaged distance function $\bar D_A(z)$ should satisfy. We will use
this equation to determine an LTB solution and the function
$\alpha(z)$, comparing its solution $\bar D_A(z)$ with a set of
observational data. 

Finally, we discuss another possibility where $\alpha(z)$ can be
smaller than $0.9$. Note that if the directions for which the  Type Ia
supernovae are observed were biased, we would  have to restrict the
integration region from the whole sphere $S^2$ to  part of it,
$\mathcal{D}_1\subset S^2$, when taking  the spherical average of the
distance equation (\ref{eq:avSachs}).  ($\mathcal{D}_1$ is not
necessarily a single connected region.) Let  $\langle f\rangle_1$ be
the partial average defined by
\begin{equation}
\langle f(z;\theta ,\phi)\rangle_1\equiv
        \frac{\int_{\mathcal{D}_1} f(z;\theta ,\phi )d\Omega}{
        \int_{\mathcal{D}_1}d\Omega}.
\end{equation}
Repeating our analysis, the averaged equation
\begin{equation}
\langle \mathcal{L}^2 D_A(z;\theta ,\phi ) \rangle_1
+4\pi G \langle \rho(z;\theta ,\phi ) 
D_A(z;\theta ,\phi )\rangle_1 =0. 
\end{equation}
becomes
\begin{equation}
\mathcal{L}_1^2 D_1(z)
+4\pi G \alpha_1(z) \rho_1(z) D_1(z) =0,
\end{equation}
where $D_1(z)\equiv\langle D_A\rangle_1$,
$\rho_1(z)\equiv\langle\rho\rangle_1$. The operator $\mathcal{L}_1^2$
is defined with respect to the partially averaged Hubble function
$\langle H(z;\theta,\phi)\rangle_1$. Function $\alpha_1(z)$ is defined
in a completely parallel way to the original $\alpha(z)$, which
implies $0.9\lesssim\alpha_1(z)\leq1.0$. To determine the
corresponding LTB spacetime, however, we need to consider the same
$\bar H(z)$ and $\bar\rho(z)$ as before with the averaging on the
whole sphere; $\bar H(z)=H_\text{LTB}(z)$ and
$\bar\rho(z)=\rho_\text{LTB}(z)$. As a result, we have
\begin{equation}
\mathcal{L}_\text{LTB}^2 D_1(z)
+4\pi G \alpha(z) \rho_\text{LTB}(z) D_1(z) =0,
\end{equation}
where
\begin{equation}
        \alpha(z)\equiv \alpha_1(z) \frac{\rho_1(z)}{\bar\rho(z)}.
\end{equation}
(We replaced $\mathcal{L}_1^2$ by $\mathcal{L}_\text{LTB}^2$ as a
rough estimate.)  This is the equation for the case of biased
distribution of observed light sources. Apparently, if the partial
average $\rho_1(z)$ was smaller than the total average $\bar\rho(z)$
the generalized $\alpha(z)$ could be smaller than $0.9$. (Actually,
any small positive value would be possible.) This may be considered as
a generalization of the idea of Dyer and Roeder
\cite{dyer:1972,dyer:1973}, which is based on a consideration of
single light ray, whereas the above equation is based on averaging
over a partial region in the sphere of constant $z$.  Since we do not
know whether the directions of SNIa are `biased' or not, in the next
section we will consider both possibilities and perform our analysis
with and without the restriction $\alpha(z)\geq 0.9$.

\section{Numerical analysis}
\label{numsol}

In this section, we numerically seek in a space of control parameters
the best configuration that can explain observational data of Type Ia
SNe without introducing dark energy, and identify the role of
anisotropy based on the results.

We employ the same void model considered by Alnes \textit{et
  al.}~\cite{Alnes:2005rw} to parametrize the LTB solution. That is,
we choose the arbitrary functions $t_B(r)$, $M(r)$ and $E(r)$ to be
\begin{align}
t_B(r) &=0, \\
M(r) &=
\frac{1}{2}H_0^2r^3\left[
  1-\frac{\Delta M}{2}  \left(
    1-\tanh{\frac{r-r_0}{2\Delta r}}\right)\right], \\
E(r) &=
\frac{1}{4}H_0^2r^2 \Delta M \left(
  1-\tanh{\frac{r-r_0}{2\Delta r}}\right),
\end{align}
where $\Delta M $ is the density contrast, $r_0$ the position of the
boundary separating the inside and the outside of the void, $\Delta r$
the width of the boundary, and $H_0$ the Hubble constant at the
center. This void model represents a flat FL universe at
$r\gg r_0$, since $M(r)\propto r^3$ and $E(r)\sim 0$ there. On the
other hand, the model represents another FL universe near the center,
since $M(r)\propto r^3$ and $E(r)\propto r^2$ there. From the latter
fact, we can define the following density parameters at the center:
\begin{align}
\Omega_{k} &\equiv
\frac{E''(0)}{H_0^2}
=\frac{\Delta M}{2} \left(1+\tanh{\frac{r_0}{4\Delta r}}\right), \\
\Omega_{m} &\equiv \frac{M'''(0)}{3H_0^2}=1-\Omega_{k}.
\end{align}
These equations allow us to use $\Omega_{m}$ to parametrize the
solution in place of $\Delta M$. We choose the ratio function
$\alpha(z)$ as constant, $\alpha(z)=\alpha=\text{constant}$, for
simplicity. Thus, our parameter space consists of $\alpha$,
$\Omega_{m}$, $r_0$, and $\Delta r$. As for the observational data to
compare, we use the so-called gold data set of Riess \textit{et al}
\cite{Riess:2004nr}, consisting of 157 samples of SNIa.

To solve the distance equation (\ref{eq:alD-R}), we need to determine
the functions $H_\text{LTB}(z)$ and $\rho_\text{LTB}(z)$ by solving
the null geodesic equations \cite{Celerier:1999hp}
\begin{align}
\frac{dt}{dz} &=-\frac{1}{1+z}\frac{R'}{\dot{R'}}, &
\frac{dr}{dz} &=\frac{1}{1+z}\frac{\sqrt{1+2E}}{\dot{R'}}, 
\end{align}
along the light path $(t(z),r(z))$
under initial conditions $t(0)=t_0$ and $r(0)=0$, with $t_0$ being the
present time. With a solution, the functions $H_\text{LTB}(z)$ and
$\rho_\text{LTB}(z)$ can be evaluated according to
$\rho_\text{LTB}(z)\equiv\rho_\text{LTB}(t(z),r(z))$ and
$H_\text{LTB}(z)\equiv H_\text{LTB}(t(z),r(z))$. 

Between the angular-diameter distance $D_A$ and the luminosity
distance $D_L$ holds in general the following relation
\cite{Etherington:1933,Ellis:1971pro}
\begin{equation}
D_L(z;\theta,\phi)=(1+z)^2D_A(z;\theta,\phi), 
\label{eq:DL-DA}
\end{equation}
which immediately implies
\begin{equation}
\bar D_L(z)=(1+z)^2 \bar D_A(z) \label{eq:DL-DAbar}.
\end{equation}
This equation allows us to convert the averaged angular-diameter
distance $\bar{D}_A$ to the averaged luminosity distance $\bar{D}_L$
to compare solutions of our distance equation with luminosity distance
data of SNIa.

To evaluate the likelihood of given parameters
$(\alpha,\Omega_m,r_0,\Delta r/r_0)$ to represent the right values, we
calculate the $\chi^2$ (e.g., \cite{Riess:1998cb})
\begin{equation}
\chi ^2(\alpha,\Omega_m,r_0,\Delta r/r_0)=
\sum_{i=1}^{157}
\left(\frac{\mu ^{\text{obs}}(z_i)-\mu (z_i)}{\sigma _i} \right)^2,
\end{equation}
where $\mu(z)$ is the distance modulus function associated with the
numerical solution $\bar D_L(z)$,
\begin{equation}
\mu(z) \equiv 5\log{(\bar D_L(z)/\text{Mpc})}+25,
\label{eq:dm}
\end{equation}
while $(z_i,\mu^{\text{obs}}(z_i),\sigma_i)$ is an observational data
in the gold data set with $z_i$ being the redshift of a supernova,
$\mu^{\text{obs}}(z_i)$ the corresponding distance modulus, $\sigma
_i$ the estimated error for $\mu ^{\text{obs}}(z_i)$. The likelihood
density is then given by
\begin{equation}
  \label{eq:P}
  P(\alpha,\Omega_m,r_0,\Delta r/r_0)=C e^{-\chi ^2/2},
\end{equation}
where $C$ is a normalization constant.

\begin{figure}[htb]
\centering
\includegraphics[width=6cm,clip]{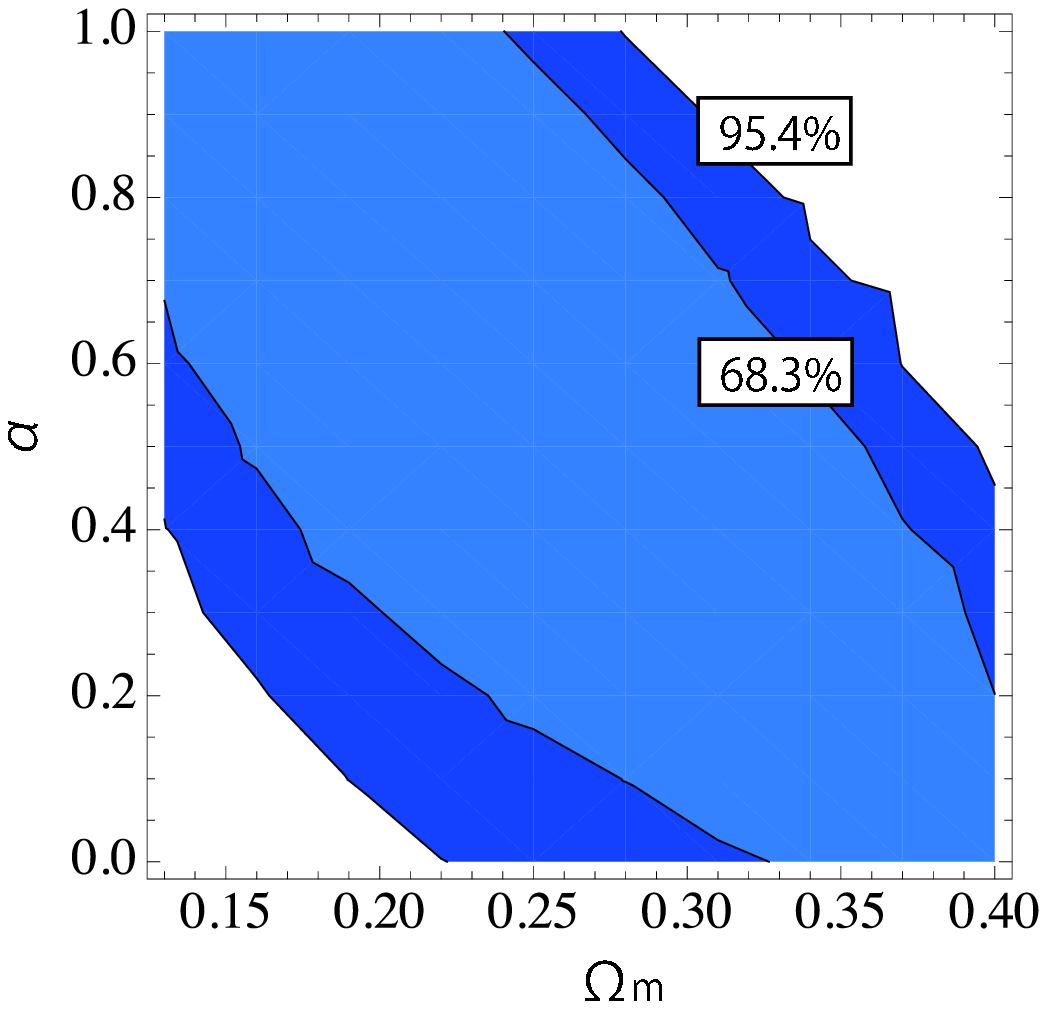}
\includegraphics[width=6cm,clip]{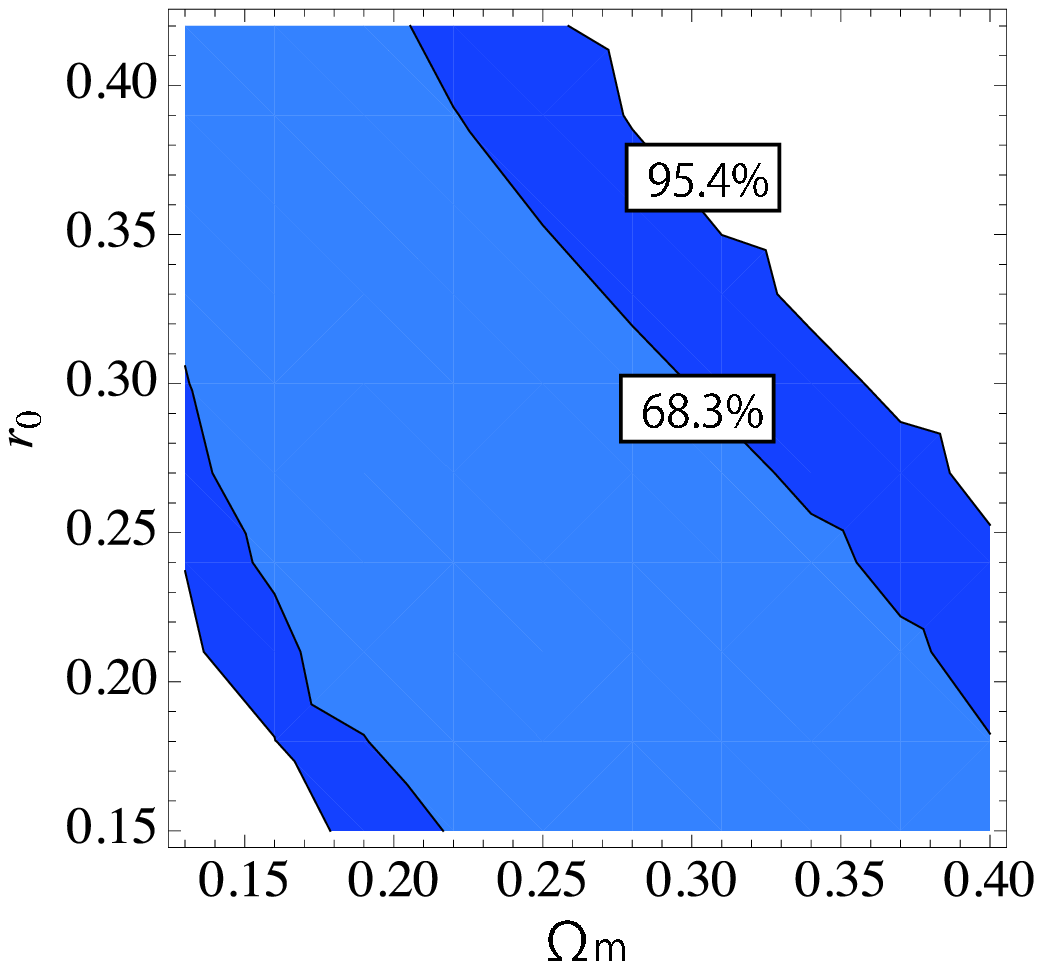}
\includegraphics[width=6cm,clip]{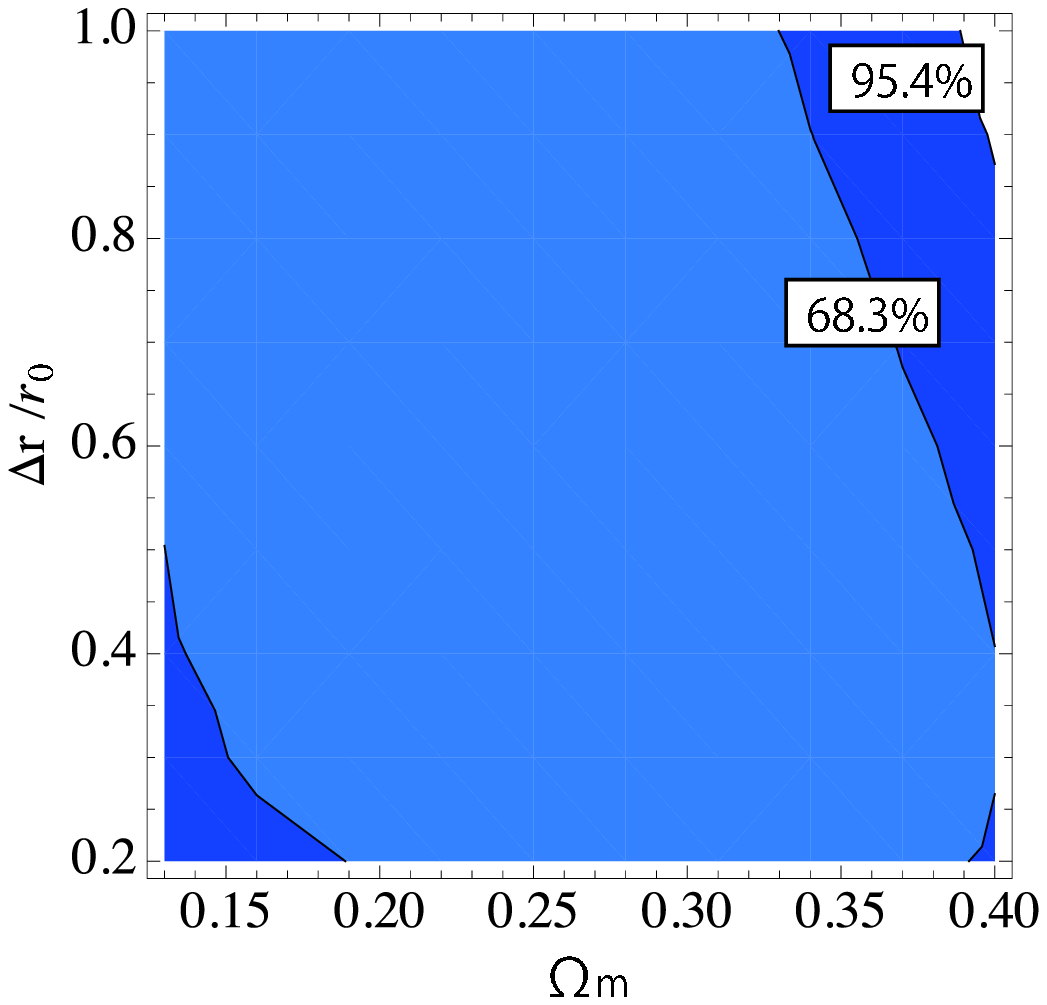}
\includegraphics[width=6cm,clip]{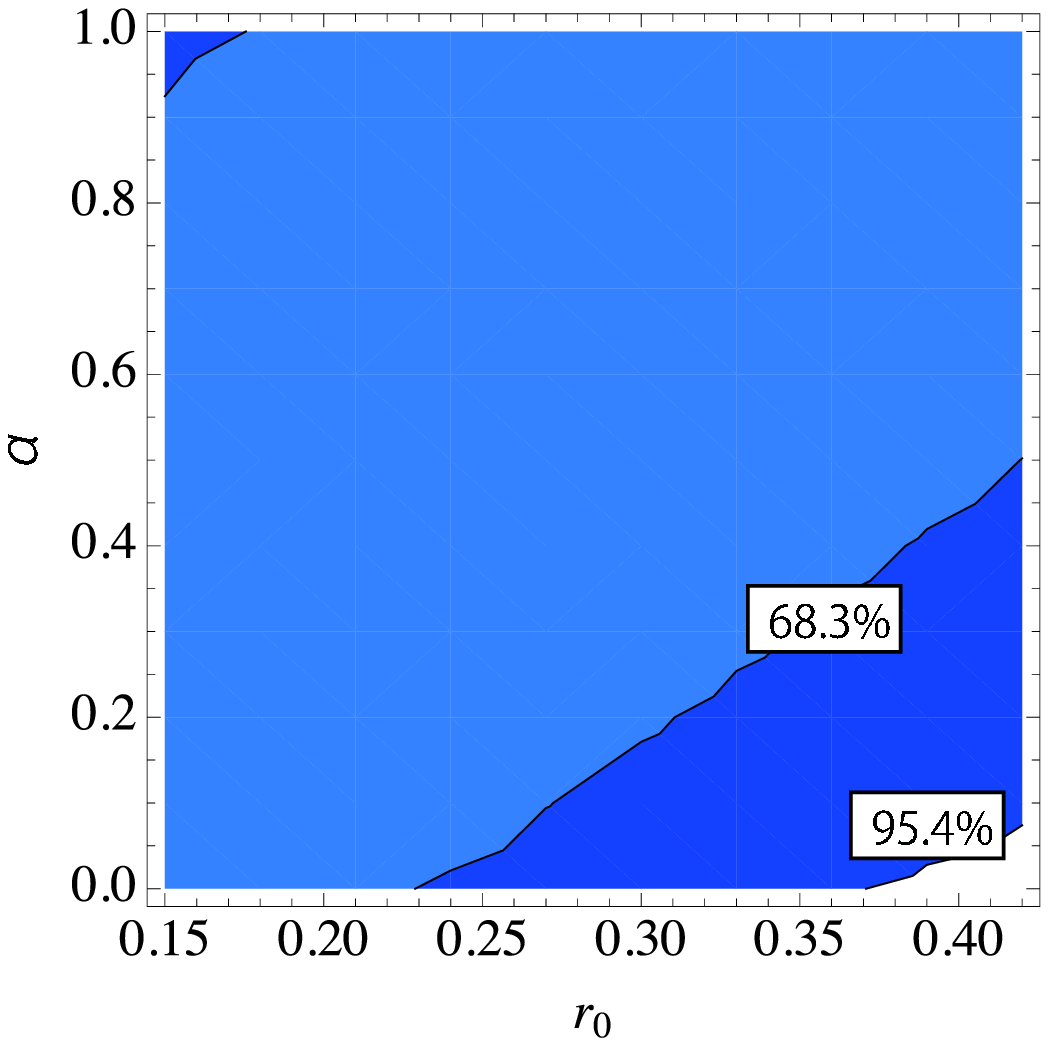}
\includegraphics[width=6cm,clip]{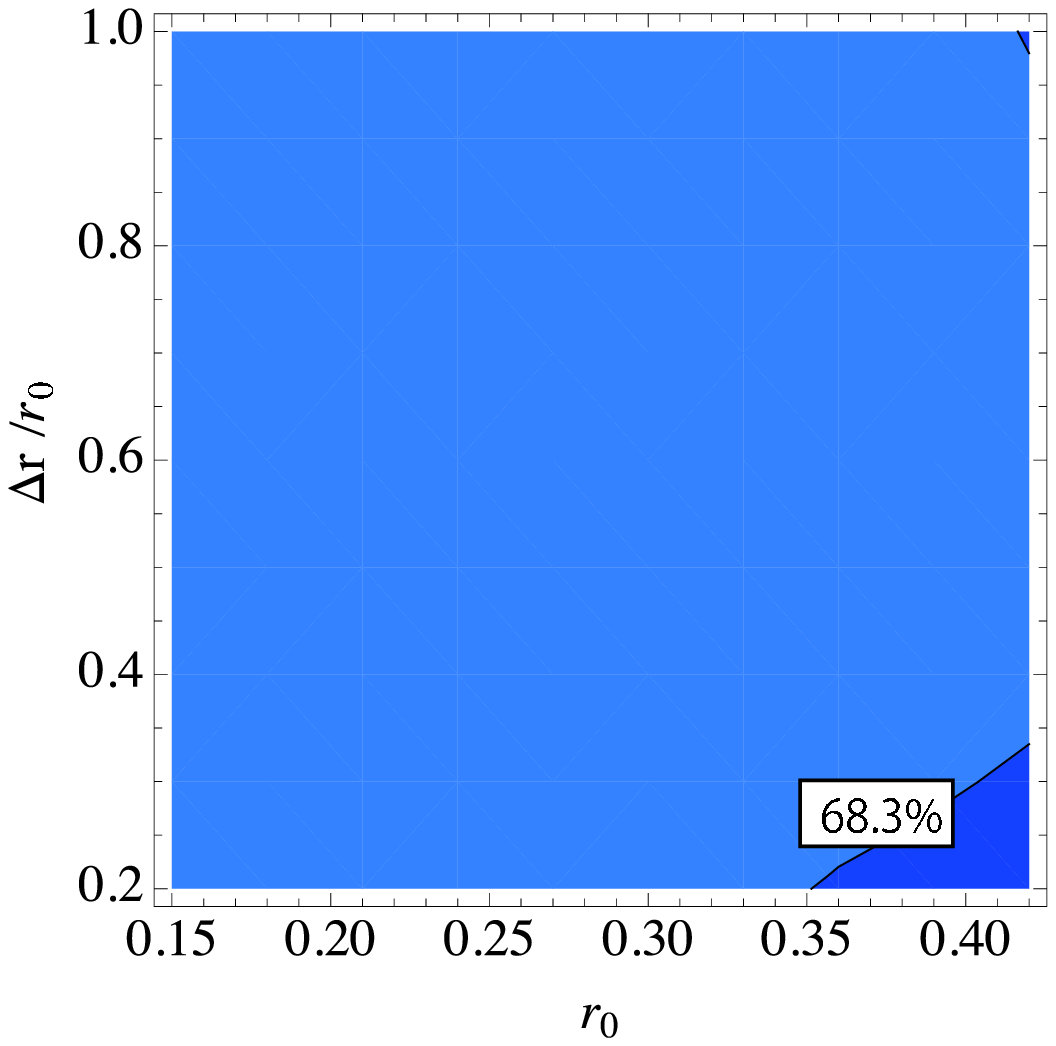}
\includegraphics[width=6cm,clip]{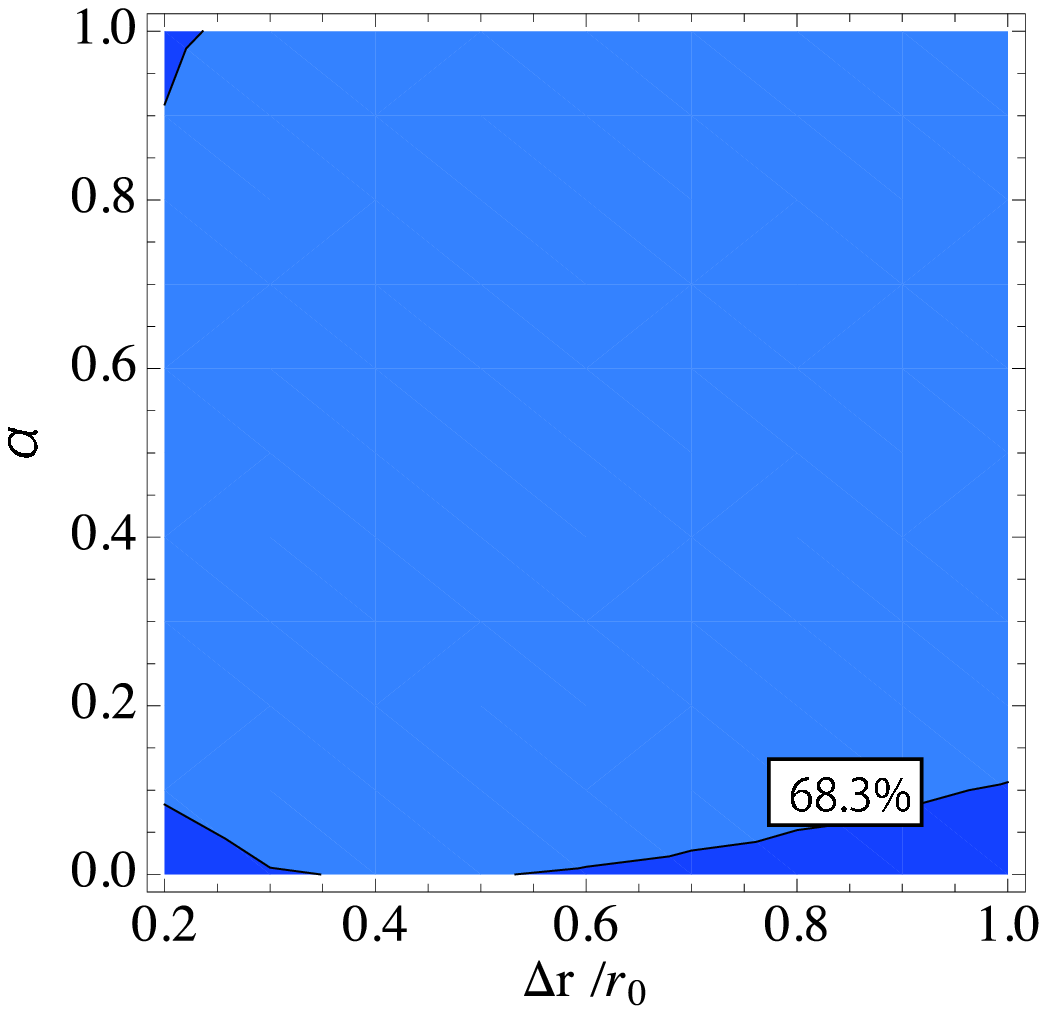}
\caption{Confidence regions in the space of two parameters obtained from
  the 4 parameter space $\mathcal{V}$, after integrating over the 
  remaining 2-dim space.}
\label{fig:contourx}
\end{figure}

\begin{figure}[htb]
\centering
\includegraphics[width=6cm,clip]{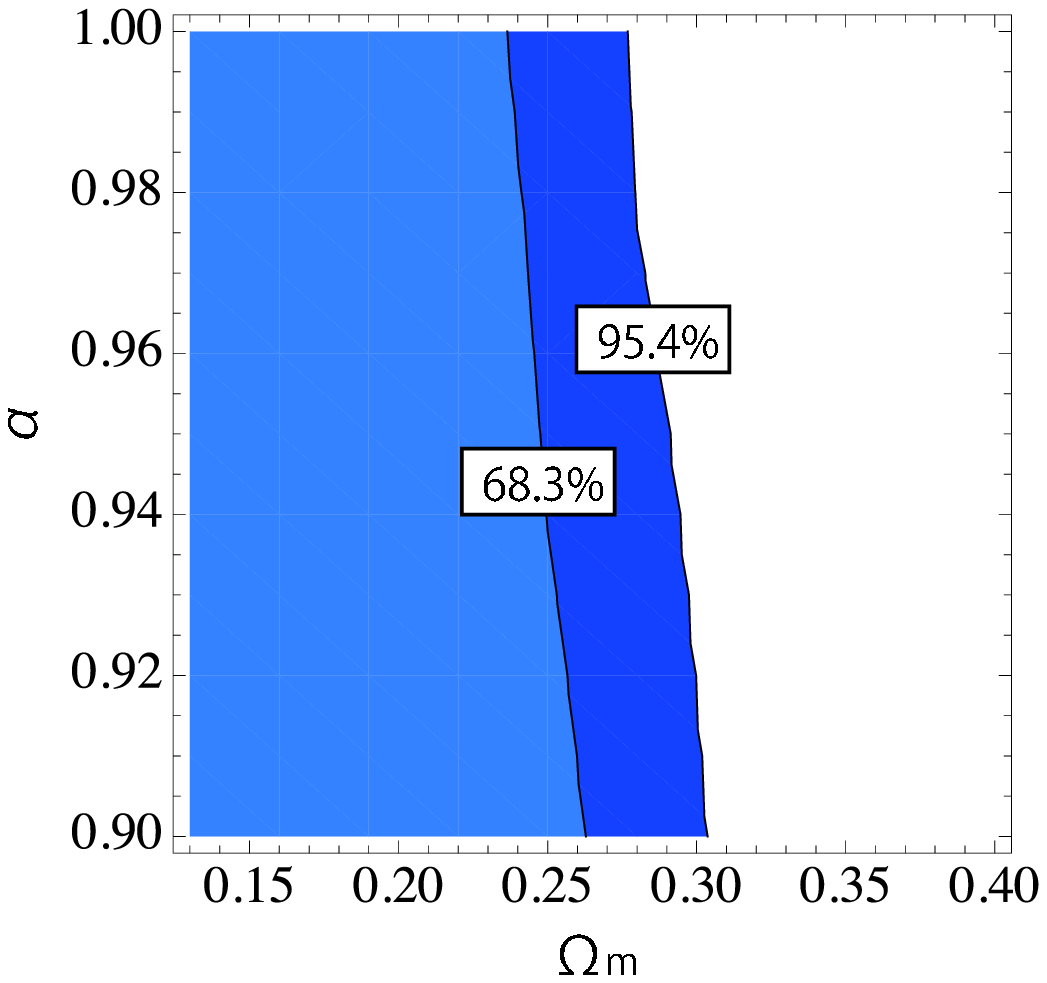}
\includegraphics[width=6cm,clip]{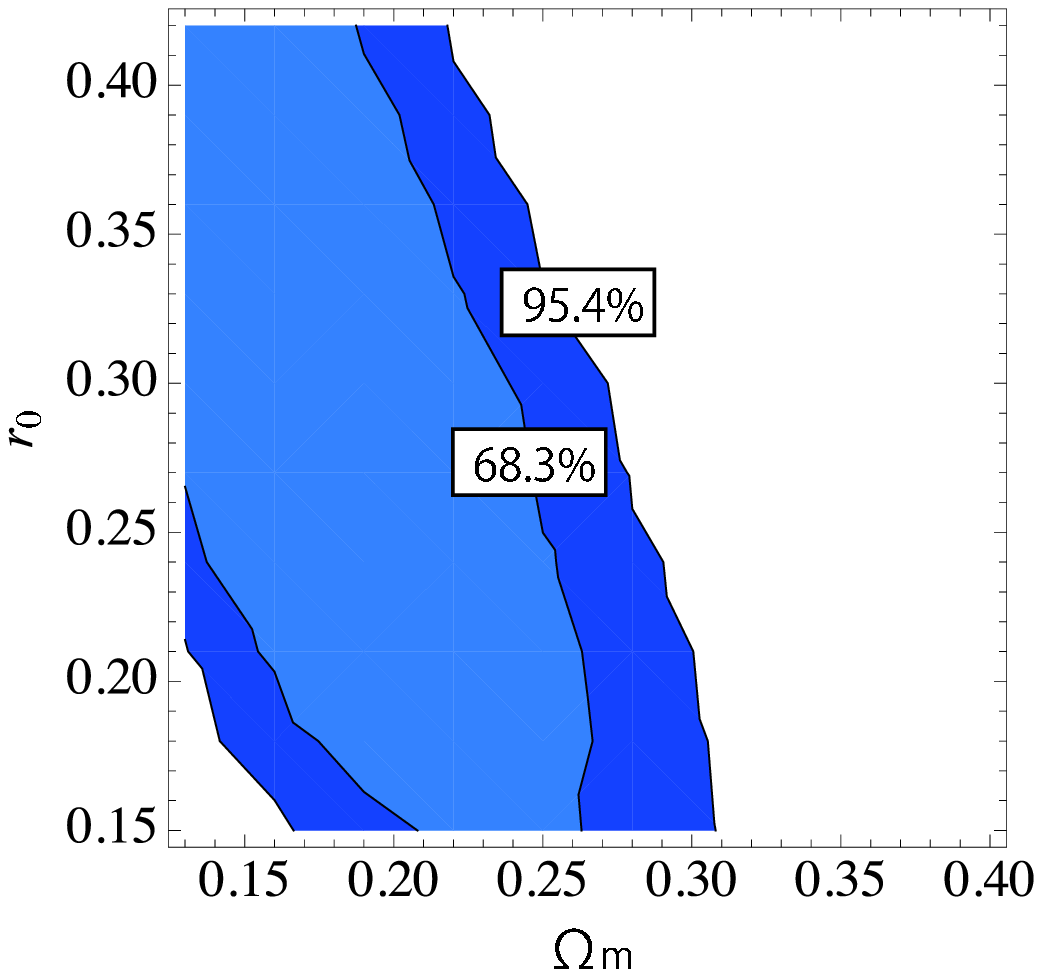}
\includegraphics[width=6cm,clip]{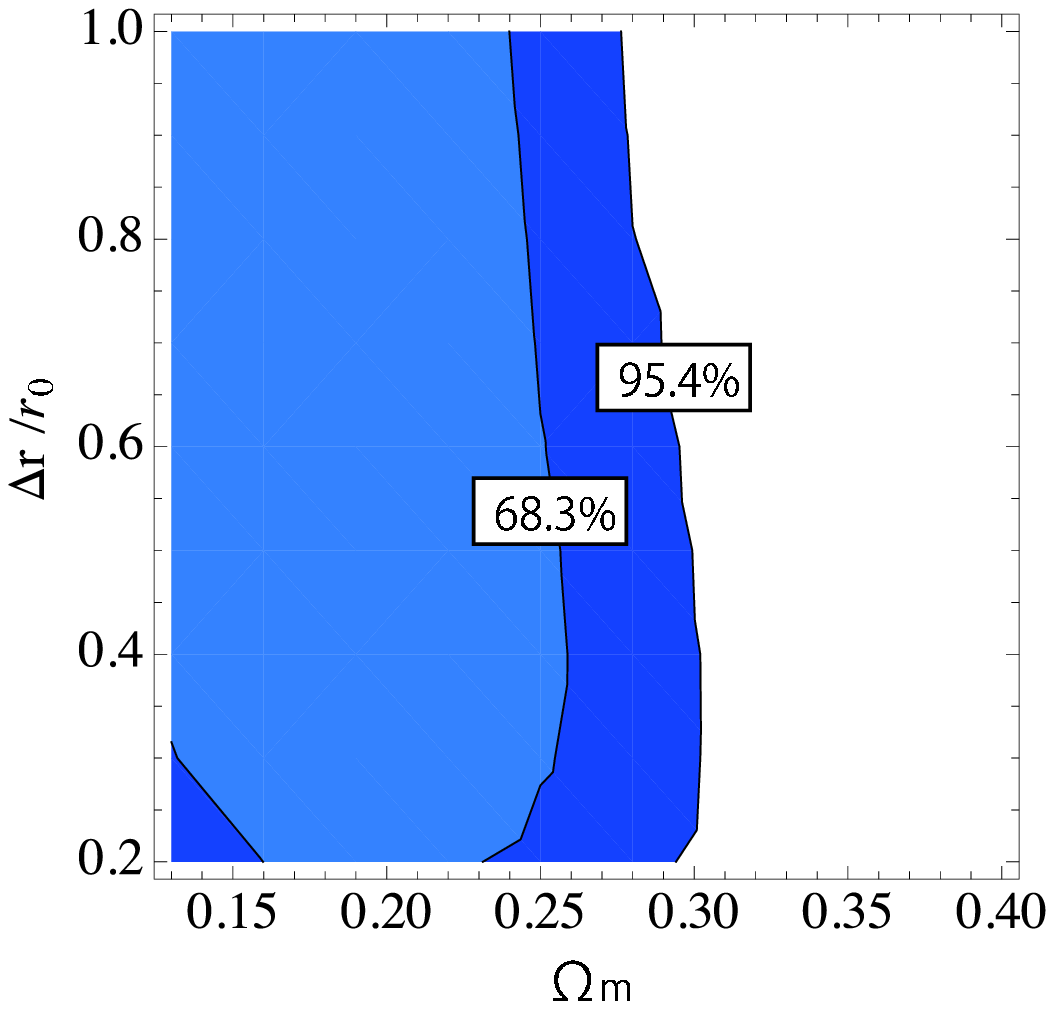}
\includegraphics[width=6cm,clip]{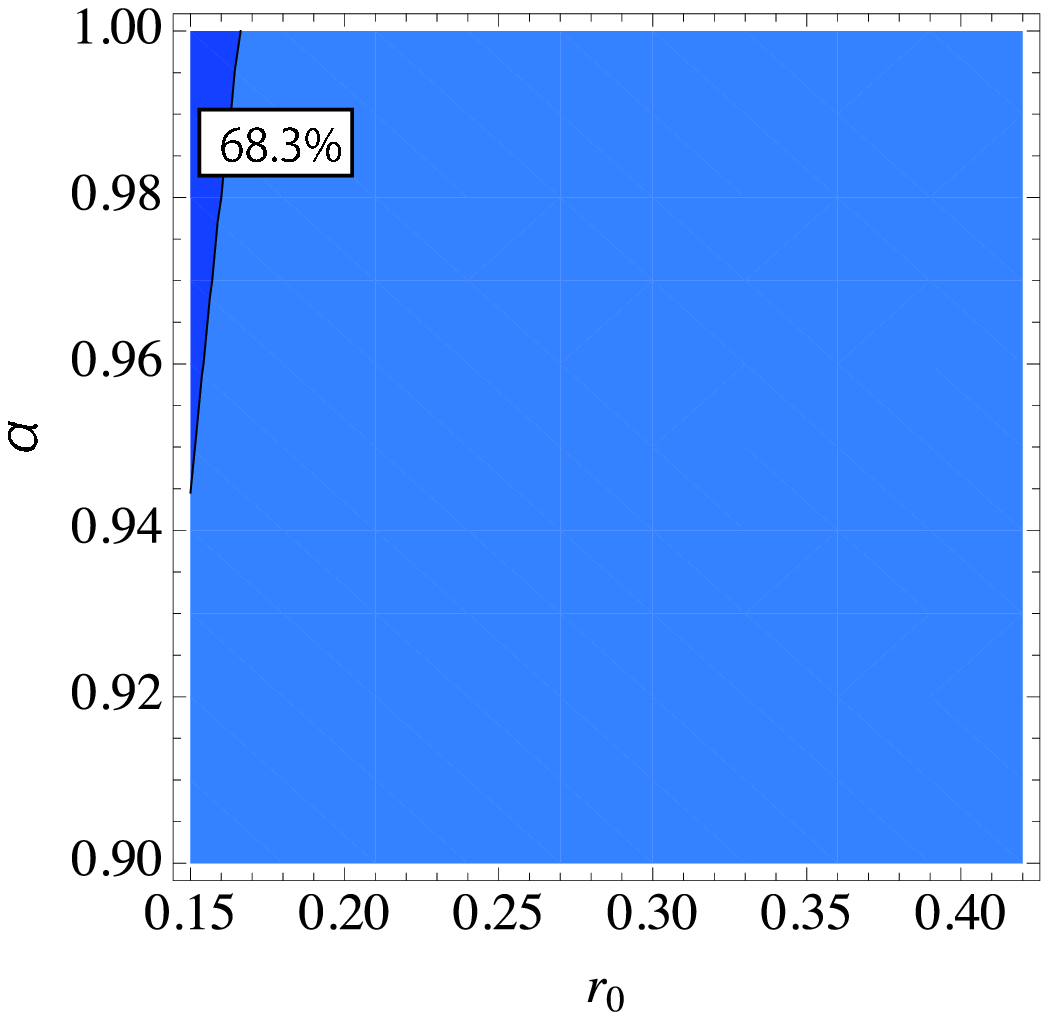}
\includegraphics[width=6cm,clip]{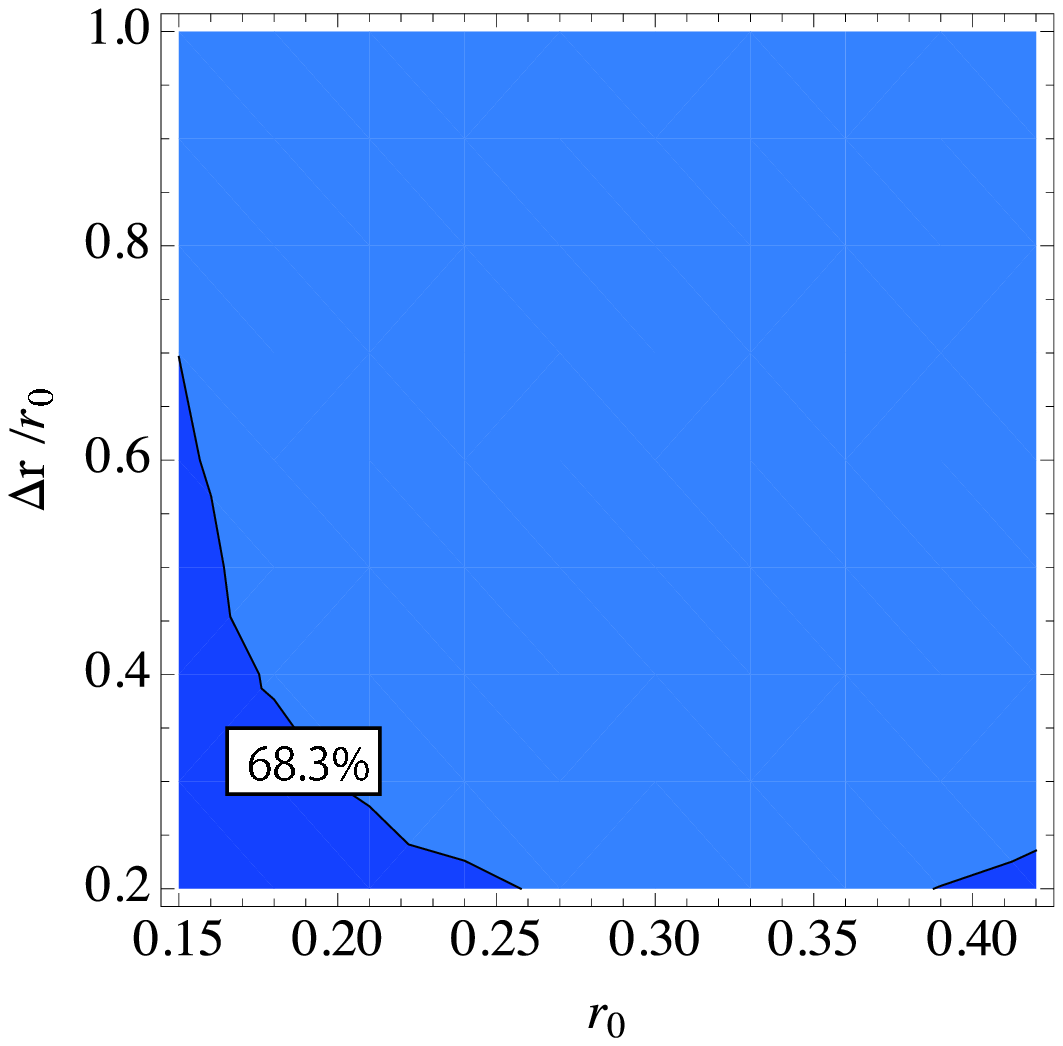}
\includegraphics[width=6cm,clip]{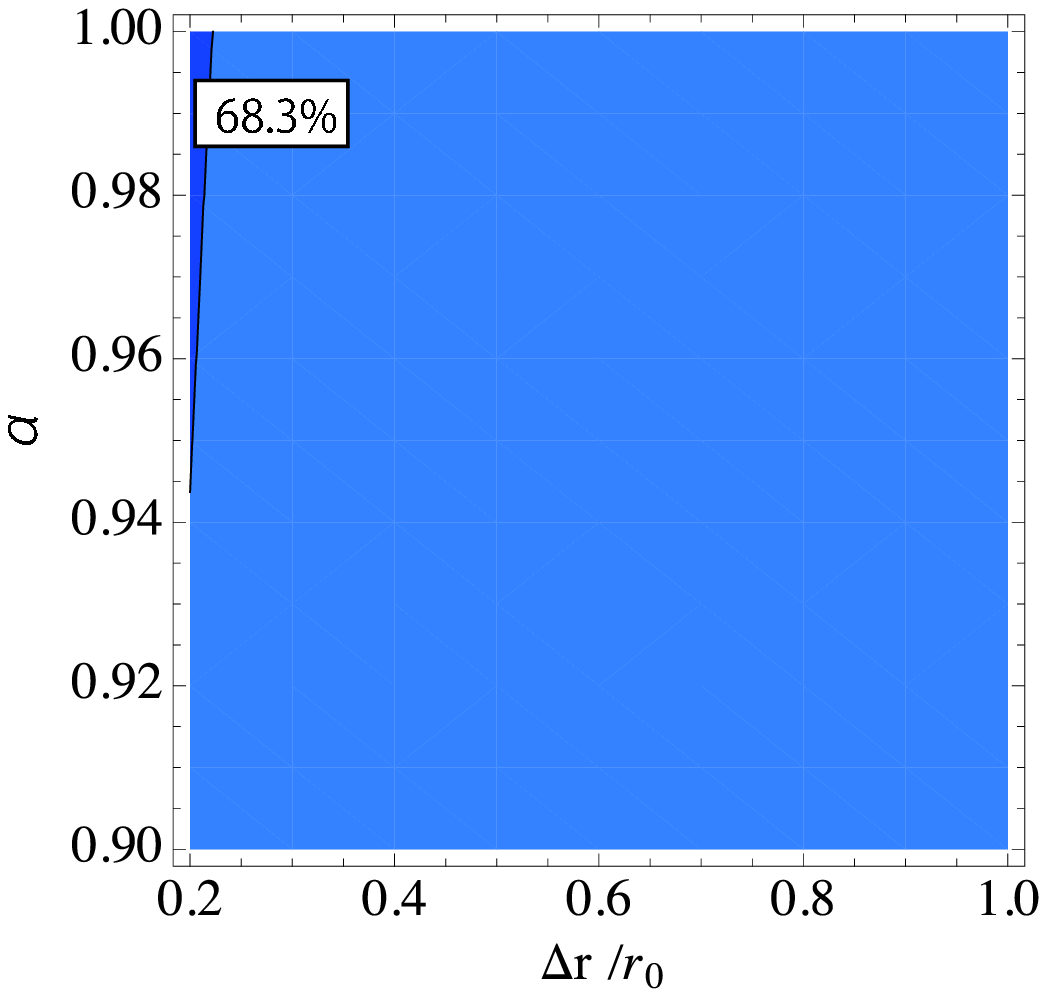}
\caption{Confidence regions in the space of two parameters obtained from
  the restricted 4 parameter space $\mathcal{V}|_{\alpha\geq0.9}$,
 after integrating over the remaining 2-dim space.}
\label{fig:contour9}
\end{figure}

\begin{figure}[htb]
\centering
\includegraphics[width=10cm,clip]{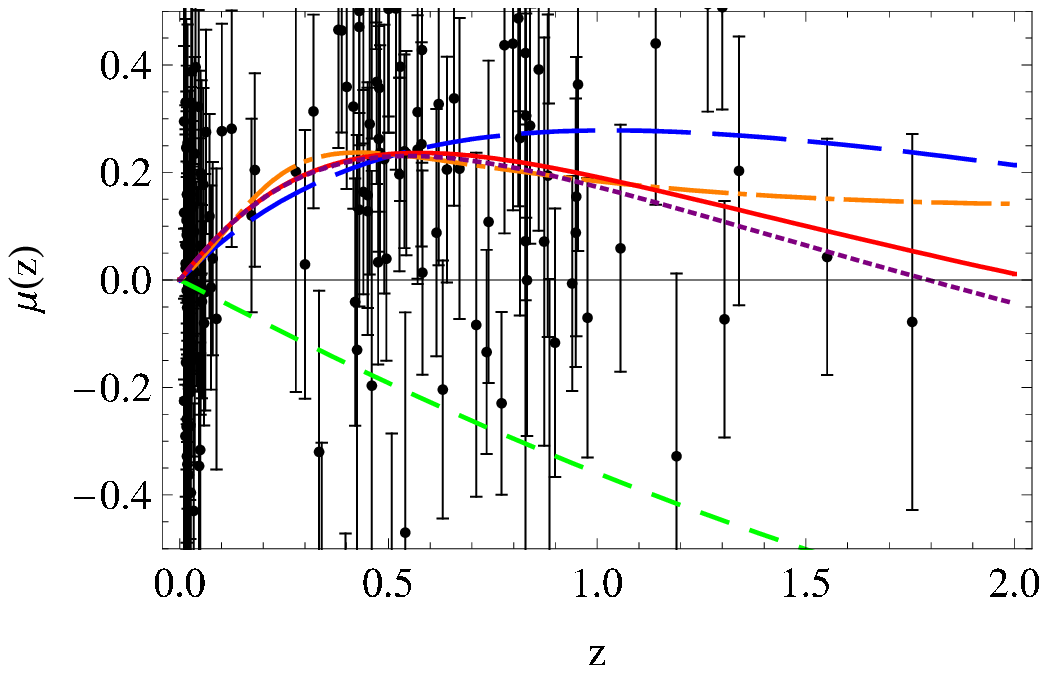}
\caption{
  Residual plots of distance modulus for the unbiased best model
  $(\Omega_m,r_0,\Delta r/r_0,\alpha)=(0.17,0.26, 0.98, 0.90)$ (red
  solid line), the biased best model $(0.31,0.18, 0.47, 0.42)$ (orange
  chain line), and the $\Lambda$CDM model
  $(\Omega_m,\Omega_\Lambda)=(0.25,0.73)$ (blue longer-interval dashed
  line) with the base horizontal line being the negative curvature FL
  model with $(\Omega_m,\Omega_k)=(0.27,0.73)$. The purple dotted
  line, corresponding to $(\Omega_m,r_0,\Delta
  r/r_0,\alpha)=(0.17,0.26, 0.98, 1.0)$, which is the same as the
  unbiased best model except the value of $\alpha$, is included to
  show how much a smaller $\alpha$ increases the distance. The
  monotonically decreasing green (shorter-interval) dashed line
  corresponds to the Einstein-de Sitter model $(\Omega_m=1)$.  The
  observational data, represented by dots and (error) bars, are taken
  from the gold samples of Riess \textit{et al.}\cite{Riess:2004nr}.  }
\label{fig:dismodralpha}
\end{figure}

We searched the parameter region defined by
\begin{equation}
  \label{eq:pdomain}
   \mathcal{V}\equiv \bigg\{(\alpha,\Omega_m,r_0,\Delta r/r_0)\bigg|\;
   \begin{matrix}
     0.0\leq \alpha\leq 1.0, &  0.13\leq\Omega_m\leq 0.40, \\
     0.15\leq r_0\leq 0.42, &  0.2\leq \Delta r/r_0\leq 1.0
   \end{matrix}
   \bigg\}.
\end{equation}
Numerically solving Eq.(\ref{eq:alD-R}) for lattice points in this
region, we sought the smallest $\chi^2$ both in whole $\mathcal{V}$
and in the region restricted to $\alpha\geq 0.9$,
$\mathcal{V}|_{\alpha\geq0.9}$.  The best-fit parameters giving the
smallest $\chi^2$ are summarized in Table \ref{tab:bestpx}. For
comparison, we included the results corresponding to the isotropic
void model $(\alpha=1)$. Figures \ref{fig:contourx} and
\ref{fig:contour9} show confidence regions in selected 2-dim parameter
spaces, obtained after integrating over the remaining 2-dim space. (We
determined the confidence regions so that the $100\%$ region
corresponds to $\mathcal{V}$ or $\mathcal{V}|_{\alpha\geq0.9}$.)
Figure \ref{fig:dismodralpha} shows residual plots.

\begin{table}[htb]
\begin{center}
\setlength{\tabcolsep}{10pt}
\footnotesize
\begin{tabular}{|c|c|c|c|} \hline
Model & Isotropic ($\alpha=1$) & Unbiased ($0.9\leq \alpha$) & 
Biased ($0\leq \alpha$) \\
\hline 
Ratio parameter: $\alpha$     & 1.0  & 0.90 & $0.42$ \\ \hline
Density parameter: $\Omega_m$ & 0.16  & 0.17 & $0.31$ \\ \hline
Transition point: $r_0$       & 0.31  & 0.26 & $0.18$ \\ \hline
Transition width: $\Delta r/r_0$ & 0.76 & 0.98 & $0.47$ \\ \hline
$\chi^2$                     & 175.0 & 174.8 & $174.1$ \\ \hline  
\end{tabular}
\end{center}
\caption{The best-fit parameters for the isotropic void model
  ($\alpha=1$), the unbiased anisotropic model ($0.9\leq \alpha\leq 1.0$),
  and the possibly biased anisotropic model ($0\leq\alpha\leq1.0$). 
\label{tab:bestpx}}
\end{table}

Our best-fit value $\chi ^2=174.1$ or $174.8$ is smaller than $\chi
^2_{\Lambda \text{CDM}}=178$ in the $\Lambda\text{CDM}$
model~\cite{Riess:2004nr} and  $\chi ^2_{\text{void}}=176.5$ in the
void model of Alnes \textit{et al.}~\cite{Alnes:2005rw}. (The value of
Alnes et al. is slightly different from our value for $\alpha=1$ shown
in Table \ref{tab:bestpx}. This is perhaps because they sought the
best parameters to fit both SNIa and the first peak of the CMB power
spectrum, while we only consider the SNIa.)

The most striking feature of $\alpha$ is that smaller value of this
parameter allows larger $\Omega_m$ and smaller $r_0$ (void
size). Linear interpolation of the values in Table \ref{tab:bestpx}
gives us the estimate
\begin{equation}
\Omega_m=-0.26\alpha+0.42, \quad r_0=0.22\alpha+0.09.
\end{equation}
The confidence regions shown in Fig.\ref{fig:contourx} also suggest
linear relations among these parameters consistent with this estimate,
although they do not exactly coincides with the above estimate. This
is because the most likely point in the whole parameter space does not
necessarily coincide with the most likely point in the reduced
parameter space that is obtained by integrating the likelihood
function over uninterested (or unfocused) parameters. As a result, the
above linear relations do not exactly coincide with the longitudinal
lines of the confidence contours in Fig.\ref{fig:contourx}. Still, the
tendencies are consistent.

This effect of $\alpha$ may be explained as follows. We know that
smaller $\alpha$ makes the luminosity distance larger (See Appendix
\ref{originalD-R}), while larger $\Omega_m$ decreases the
distance. Smaller void size also decreases the distance, since our
void is defined as an underdense region, where smaller (possibly
negative) curvature has an effect of increasing distance. Therefore
smaller $\alpha$ can compensate larger $\Omega_m$ and smaller void
size.  Our numerical results seem to confirm this explanation.

The parameter $\Delta r/r_0$, which is the boundary width of the void
(relative to its radius), appears to be rather insensitive to the
likelihood; the confidence regions for $\Delta r/r_0$ versus
$\Omega_m$, $r_0$, or $\alpha$ shown in Fig.\ref{fig:contourx} do not
provide reasonably sharp prediction about the choice of $\Delta
r/r_0$. This parameter may therefore be considered less important.

\section{Concluding remarks} \label{conclusion}

Since Type Ia supernovae are observed in various directions in the
sky, the distance-redshift relation inferred by a collective data of
them should be cosidered as an average over the solid angle. On the
other hand, the distance-redshift relation (distance function)
$D_A(z)$ (or $D_L(z)$) for a given direction of sight can be
theoretically determined by the distance equation that is equivalent
to the optical scalar equation. To obtain an effective equation for
the spherically averaged distance function $\bar D_A(z)$ (or $\bar
D_L(z)$), with which the observed distance-redshift relation can
directly be compared, we took an average of the distance equation, and
found that the effective equation we want is given by the distance
equation for the LTB universe with a Dyer-Roeder-like extension.  The
function $\alpha(z)$ introduced in our effective equation represents
the degree to which the Universe fluctuates from an isotropic
configuration. Smaller $\alpha$ generally implies larger fluctuation
of matter. Our numerical computations demonstrated that smaller
$\alpha$ allows larger $\Omega_m$ and smaller void size. This will
provide a useful guideline for further investigations of inhomogeneous
models.

In our numerical study, we chose the function $\alpha(z)$ to be
constant for simplicity, but in reality this function should be a
monotonically increasing function such that $\alpha(z)\rightarrow1
(z\rightarrow\infty)$, reflecting the fact that the Universe was
sufficiently isotropic (and homogeneous) at the last scattering
surface. In this paper we only considered luminosity distance-redshift
data of SNIa as obserbational data to compare; for this purpose
constant $\alpha$ was sufficient. However, to include other kind of
data like CMB power spectrum it is inevitable to have a varying
$\alpha$, resulting in more parameters to search. It would still be
worth exploring such larger parameter space and compare with all
available obserbational data including CMB and that of baryon acoustic
oscillation. This is a future work.

\appendix
\section{The original Dyer-Roeder extension} \label{originalD-R}

Dyer and Roeder \cite{dyer:1972,dyer:1973} considered a simplified
model of the Universe in which the matter is almost concentrated into
objects such as galaxies, and the light from a distant object travels
through the intergalactic, almost empty space. They assume that the
geometry is described by the usual Robertson-Walker metric
\begin{equation}
ds^2 = -dt^2+a(t)^2 \left[
  \frac{dr^2}{1-Kr^2}+r^2(d\theta ^2+\sin^2{\theta }d\phi ^2) \right],
\end{equation}
where $a(t)$ is the scale factor and $K$ is the curvature index.

Let $(t(\lambda),r(\lambda))$ be a null geodesic coming to an
observer, with $\lambda$ being the affine parameter. Then, the
geodesic equation for $t(\lambda)$ and the null condition in the FL
universe are
\begin{align}
\frac{d^2t}{d\lambda^2} &=\frac{-a\dot{a}}{1-Kr^2}\left(
  \frac{dr}{d\lambda}\right)^2, &
-\left(\frac{dt}{d\lambda } \right)^2 +\frac{a^2}{1-Kr^2}\left(
  \frac{dr}{d\lambda } \right)^2 &=0,
\end{align}
respectively.
Using these equations, Eq.(\ref{eq:lam-z}) yields 
\begin{align}
\frac{d}{d\lambda } &=\frac{1}{\omega(0)}\frac{\dot{a}}{a}\left(
  \frac{dt}{d\lambda } \right)^2\frac{d}{dz} \notag \\
&=\omega(0)\frac{\dot{a}}{a}(1+z)^2\frac{d}{dz} \notag \\
&=\omega(0)H_{\text{FL}}(1+z)^2(z)\frac{d}{dz}, \label{eq:l-zFLRW}
\end{align}
where we have used $z=\frac{-1}{\omega(0)}\frac{dt}{d\lambda }-1$ in
the second equality, and $H_{\text{FL}}(z)\equiv \dot{a}/a$ is the
Habble parameter for the FL universe.  With Eq.(\ref{eq:l-zFLRW}),
this yields the distance equation
\begin{equation}
\mathcal{L}_\text{FL}D^{(\text{FL})}_A(z)+
4\pi G\rho_{\text{FL}}(z) D^{(\text{FL})}_A(z)=0,
\label{eq:D-RFLRWDA}
\end{equation}
where we have defined
\begin{equation}
  \mathcal{L}_\text{FL}\equiv H_{\text{FL}}(z)\frac{d}{dz}\left[
    H_{\text{FL}}(z)(1+z)^2\frac{d}{dz} \right],
\end{equation}
and $D^{(\text{FL})}_A$ is the angular-diameter distance in the FL
universe and $\rho_{\text{FL}}$ the matter density.

To take into account the effect of lumps, Dyer and Roeder reduced, in
the above equation, the matter density $\rho_{\text{FL}}(z)$ on the
light path to $\alpha_\text{DR} \rho_{\text{FL}}$ by a constant
$\alpha_\text{DR}(0\leq \alpha_\text{DR} \leq 1)$. As a result, the
Dyer-Roeder equation is given by
\begin{equation}
\mathcal{L}_\text{FL} D^{(\text{DR})}_A(z)
+4\pi G\alpha_\text{DR} \rho_{\text{FL}}(z) D^{(\text{DR})}_A(z)=0,
\label{eq:D-RFLRWal}
\end{equation} 
where $D^{(\text{DR})}_A(z)$ is the angular-diameter distance in a
lumpy universe with parameter $\alpha_\text{DR}$. In this equation,
the value $\alpha_\text{DR} =1$ corresponds to the homogeneous
universe, and the value $\alpha_\text{DR} =0$ to a universe where all
the matter is concentrated into lumps.

If the universe is flat ($K=0$), we can easily solve
Eq.(\ref{eq:D-RFLRWal}) under the initial conditions
$D^{(\text{DR})}_A(0)=0$ and $dD^{(\text{DR})}_A(0)/dz=1/H_0\equiv
1/H_\text{FL}(0)$.  The solution is
\begin{align}
D^{(\text{DR})}_A(z) &=\frac{2}{H_0 \sqrt{25-24\alpha_\text{DR} }}\left[
  (1+z)^{-b_1}-(1+z)^{-b_2} \right ],  \\
 \bigl( b_1 &\equiv \frac{5-\sqrt{25-24\alpha_\text{DR} }}{4},
 b_2 \equiv \frac{5+\sqrt{25-24\alpha_\text{DR} }}{4} \bigr).
\end{align}
From Eq.(\ref{eq:DL-DA}), the corresponding luminosity distance is
\begin{equation}
D^{(\text{DR})}_L(z) =
\frac{2(1+z)^2}{H_0 \sqrt{25-24\alpha_\text{DR}}}\left[
  (1+z)^{-b_1}-(1+z)^{-b_2} \right ]. 
\end{equation}

\begin{figure}
\centering
\includegraphics[width=10cm,clip]{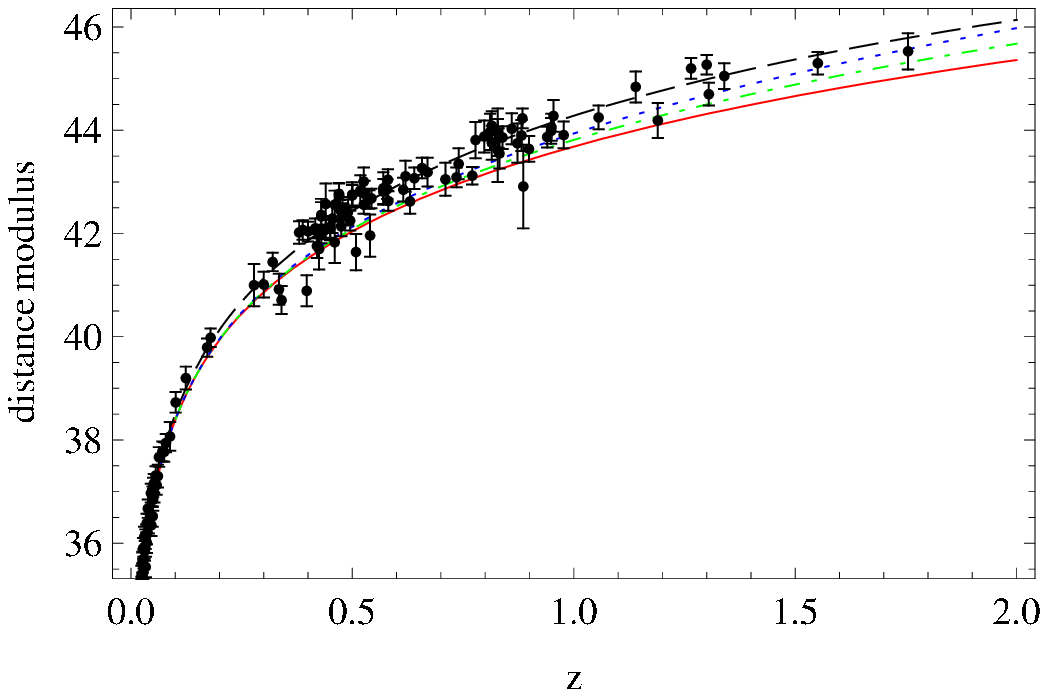}
\caption{The $\mu$-$z$ relations in the original Dyer-Roeder model
  with observational data.  The solid (red), chain (green), and dotted
  (blue) lines correspond to $\alpha_\text{DR}=1$, $\alpha_\text{DR}
  =0.5$, and $\alpha_\text{DR} =0$, respectively. The dashed (black)
  line corresponds to the $\Lambda$CDM model. The observational data
  are the gold data\cite{Riess:2004nr}.}
\label{fig:FLRWmzplot}
\end{figure}

Fig.\ref{fig:FLRWmzplot} shows the relation between the distance
modulus Eq.(\ref{eq:dm}) and the redshift $z$ for various
$\alpha_\text{DR}$. As seen from these plots, the distance increases
when $\alpha <1$ as compared to $\alpha =1$.  (Actually we can show
that the distance monotonically increases as $\alpha_\text{DR}$
decreases, i.e., $\partial D_A(z;\alpha_\text{DR})/\partial
\alpha_\text{DR}<0$, for this case.) In other words, the distance
takes the maximum at $\alpha_\text{DR}=0$, but it still insufficient
to explain the observation.

\end{document}